\documentclass[%
 aip,
 amsmath,amssymb,
 reprint,%
]{revtex4-1}

\usepackage{graphicx}
\usepackage{dcolumn}
\usepackage{bm}

\usepackage{xcolor}

\usepackage[utf8]{inputenc}
\usepackage[T1]{fontenc}
\usepackage{mathptmx}
\usepackage{etoolbox}

\makeatletter
\def\@email#1#2{%
 \endgroup
 \patchcmd{\titleblock@produce}
  {\frontmatter@RRAPformat}
  {\frontmatter@RRAPformat{\produce@RRAP{*#1\href{mailto:#2}{#2}}}\frontmatter@RRAPformat}
  {}{}
}%
\makeatother
\begin{document}


\title{Temperature crossovers in the specific heat of amorphous magnets}
\author{H\'ector Ochoa}
 \email{ho2273@columbia.edu}
\affiliation{Department of Physics, Columbia University, New York, NY 10027, USA}

\date{\today}

\begin{abstract}
It is argued that the specific heat of amorphous solids at low temperatures can be understood to arise from a single branch of collective modes. The idea is illustrated in a model of a \textit{correlated spin glass} for which magnetic anisotropies are present but they are completely frustrated by disorder. The low-energy spectrum is dominated by soft modes corresponding to propagating Halperin-Saslow spin waves at short wavelengths, evolving into a relaxation and a diffuson mode at the long wavelengths. The latter gives rise to an anomalous temperature behavior of the specific heat at $T\ll T_{*}$: $C\sim T$ in $d=3$ solids, $C\sim T\ln(1/T)$ in $d=2$. The temperature scale $T_{*}$ has a non-trivial dependence on the damping coefficient describing magnetic friction, which can be related to fluctuations of the spin-torque operator via a Green-Kubo formula. Halperin-Saslow modes can also become diffusive due to disclination motion (plastic flow). Extensions of these ideas to other systems (including disordered phases of correlated electrons in cuprate superconductors and of moir\'e superlattices) are discussed.
\end{abstract}

\maketitle

\section{Introduction}

The central theme of this article is an old question relative to the low-temperature properties of amorphous (or more generically, disordered or \textit{glass}) materials and their apparent excess of boson modes at low energies. Specifically, the question addressed here was directly posed by Brian Coles to Philip Anderson (different accounts point to each of them for coining the term \textit{spin glass}) about 50 years ago:\cite{Anderson}
\begin{quotation}
\textit{If one were to start from a totally different view and regard the spin glasses and real glasses as not too bad a deviation from a long range ferromagnet or antiferromagnet and a long range lattice, and say what we have are excitations or damped spin waves on the one hand and damped phonons on the other which would distort our spin wave specific heat, could you possibly get out the linear specific heat?}
\end{quotation}
Coles' different view expressed in this question was in contrast to the tunnelling two-level states (TTLS) of Anderson, Halperin, Varma\cite{Anderson_Halperin_Varma} and, independently, of Phillips,\cite{Phillips} which Anderson had just presented in the same conference. Anderson's answer was, according to the same record (here abbreviated):\begin{quotation}
\textit{That doesn't work because the phonon specific heat is visible and it is $T^3$. On the other hand, one of the most fascinating things about the ordinary glasses is that the coefficient of $T^3$ is wrong, relative to sound velocity.}
\end{quotation}
Coles insisted:\begin{quotation}
\textit{Doesn't that mean that part of the quasi phonon like excitations are no longer going as $T^3$ but as $T$?}
\end{quotation}
To what Anderson finally replied: \begin{quotation}
\textit{Well, the point is you see the phonons, in fact, you see more than you expect from the phonons already and that is something else again. That is really there as $T^3$. There are no theorems, but I have a feeling (I think Brenig has made this point also) that there are in some sense two entirely separate branches to the spectrum. There is a long wave length collective branch, but there is at the same time something else at very short wave lengths that is very local.}
\end{quotation}
Anderson's final point is certainly valid in many situations. The TTLS picture gained quickly acceptance thanks to various experimental observations, in particular, the dependence of the specific heat on the experimental observation times and the saturation of acoustic attenuation at high intensities.\cite{Phillips2}

However, these observations do not imply that the TTLS model is necessarily of universal application. 
An eloquent critique in that regard has been expressed by Leggett and Vural.\cite{Leggett_Vural} More broadly, the appearance of these short-wavelength excitations at low energies, specifically, the finite density of states at zero energy (constant in the original model) is the consequence of the chosen distribution of model parameters of the two-level systems. This choice is \textit{ad hoc} and not easily justified from microscopics. For example, Anderson \textit{et al}. originally wrote:\cite{Anderson_Halperin_Varma} 
\begin{quotation}
\textit{It is not easy to speculate meaningfully on the distribution of the barriers. Most of the simple models we have tried tend to suggest a predominance of rather high barriers; but the observations, as discussed above, suggest rather that there is a very broad distribution of barrier heights, with low ones as a probable as high ones.}
\end{quotation}
Here the \textit{barriers} is one of the TTLS model parameters: the energy barrier separating two minima in the complex energy landscape of the amorphous material. Since the model is heuristic, it cannot provide us with an estimate of the temperature scale corresponding to a crossover from a TTLS- to a phonon-dominated specific heat. This is the main goal of this manuscript starting from Coles' point of view instead. 

This point of view has been recently put forward by Baggioli and Zaccone\cite{BZ1,BZ2} borrowing intuition in part from incommensurate lattices,\cite{incom1,incom2,incom3} where a phason can play the role of a(n) (over-)damped phonon to give a linear-in-$T$ specific heat.\cite{Cano_Levanyuk} In this article, I elaborate on these ideas trying to establish a connection with microscopic models. I present a discussion of the long-time dynamics of a spin model for amorphous magnets forming a so-called \textit{correlated spin glass}.\cite{CSG} At short wavelengths, the hydrodynamic modes of this phase resemble propagating Halperin-Saslow spin waves.\cite{Halperin_Saslow} At long wavelengths, one mode is fully relaxed and the other becomes diffusive. In this model the mechanism is ultimately related to the lack of a local conservation law for spin angular momentum. Alternatively, in an Edwards-Anderson spin glass,\cite{Edwards-Anderson} Halperin-Saslow modes become diffusive due to the dissipative motion of vortex disclinations of the order pararmeter.\cite{DV1,DV2} Associated with the change in character of the soft modes there are crossovers in the temperature behavior of various physical properties, in particular, in the spin wave-dominated specific heat. 
The discussion is extended to other disordered systems.

\section{Two models of a spin glass}

\begin{figure}
\includegraphics[width=\columnwidth]{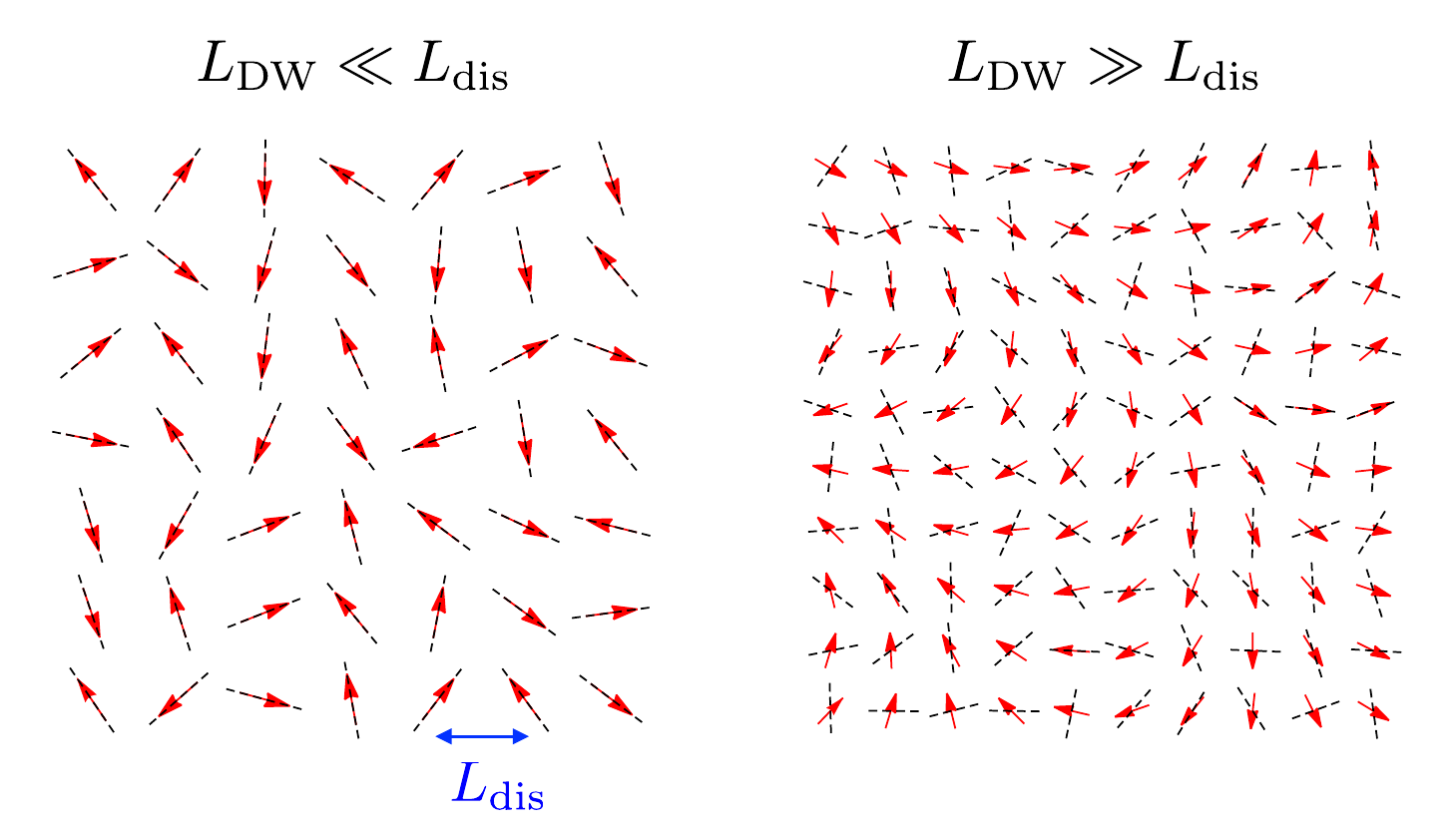}
\caption{\label{fig:sketch} Possible spin textures in the ground state of the model of Eq.~\eqref{eq:model2}. The second case corresponds to a \textit{correlated spin glass}.\cite{CSG}}
\end{figure}

Textbook models of a spin glass are usually described by a Hamiltonian of the Heisenberg type,\begin{align}
\label{eq:model1}
\hat{H}=-\frac{1}{2}\sum_{ij} J_{ij} \hat{\boldsymbol{S}}_i\cdot \hat{\boldsymbol{S}}_j,
\end{align}
where $\hat{\boldsymbol{S}}_i$ are spin operators satisfying $[\hat{S}_i^{\alpha},\hat{S}_j^{\beta}]=i\epsilon_{\alpha\beta\gamma}\hat{S}_{i}^{\gamma}\delta_{ij}$ defined on a lattice of $N$ sites, not necessarily periodic, and $J_{ij}$ are exchange coupling constants between sites $i$ and $j$; $\epsilon_{\alpha\beta\gamma}$ is the Levi-Civita symbol and repeated greek indices are summed hereafter. In a spin glass, the couplings $J_{ij}$ are treated as random variables with zero mean. Collinear magnetic order with a net magnetization is precluded, yet individual spins can freeze below certain temperature to form a complex non-collinear spin texture. This form of order is described by a finite value $q$ of an Edwards-Anderson order parameter,\cite{Edwards-Anderson}\begin{align}
q\delta_{\alpha,\beta}=\frac{1}{N}\sum_i  \langle \hat{S}_i^{\alpha} \rangle \langle \hat{S}_i^{\beta} \rangle,
\end{align}
where $ \langle \cdot \rangle$ denotes a statistical averages (see concrete definition below).

Alternatively, we may consider a different model to describe amorphous magnets (e.g. in the case of rare-earth transition-metal compounds) given by the Hamiltonian\cite{Harris}\begin{align}
\label{eq:model2}
\hat{H}=-J\sum_{\left\langle ij\right\rangle} \hat{\boldsymbol{S}}_i\cdot \hat{\boldsymbol{S}}_j-K\sum_i\left[\boldsymbol{\xi}_i\cdot\hat{\boldsymbol{S}}_i\right]^2,
\end{align}
where exchange interactions are restricted to nearest neighbors in a lattice of period $a$, and $K$ represents an easy-axis magnetic anisotropy imposed by spin-orbit coupling and the crystalline environment. In a polycrystalline material, the anisotropy axis $\boldsymbol{\xi}_i$ can be taken as a random variable with zero mean (indicating no global preferred direction of anisotropy) and correlated over a length $L_{\textrm{dis}}$.

These models differ in two crucial aspects: i) The Hamiltonian in Eq.~\eqref{eq:model1} is invariant under spin rotations; the Hamiltonian in Eq.~\eqref{eq:model2} is not. ii) In the model described by Eq.~\eqref{eq:model1} there is a single microscopic energy/length scale; in the model of Eq.~\eqref{eq:model2}, there are actually two, the length $L_{\textrm{dis}}$ introduced by disorder correlations in the $\boldsymbol{\xi}_i$-orientations, and the intrinsic length scale resulting from the competition between the two terms in the Hamiltonian, $L_{\textrm{DW}}=a\sqrt{J/K}$. In a crystalline material, this length can be interpreted as the characteristic width of a domain wall connecting spin configurations parallel and anti-parallel to a fixed anisotropy axis. In a polycrystalline material, this length together with a finite $L_{\textrm{dis}}$ (think of it as a the crystalline grain size) defines a third mesoscopic scale corresponding to the magnetic correlation length in the sense of Imry-Ma,\cite{Imry-Ma}\begin{align}
\label{eq:Lc}
L_{\textrm{c}}\approx L_{\textrm{dis}}\left(\frac{L_{\textrm{DW}}}{L_{\textrm{dis}}}\right)^{\frac{4}{4-d}},
\end{align}
where $d=2,3$ is the dimensionality of the system.

This last equation describes two different types of spin textures (see Fig.~\ref{fig:sketch}) below the freezing temperature arising from the competition between $L_{\textrm{dis}}$ and $L_{\textrm{DW}}$, or equivalently, the two energy scales in the Hamiltonian of Eq.~\eqref{eq:model2}. If $L_{\textrm{DW}}\ll L_{\textrm{dis}}$, anisotropy dominates over exchange and one should expect a texture in which the magnetization follows closely the local anisotropy direction; the magnetic correlation length saturates to $L_{\textrm{c}}=L_{\textrm{dis}}$. In the opposite limit, $L_{\textrm{DW}}\gg L_{\textrm{dis}}$, exchange dominates over anisotropy, producing a smooth spin texture varying on the scale $L_{\textrm{c}}\gg L_{\textrm{dis}}$. This is known as a \textit{correlated spin glass}.\cite{CSG}


\section{Hydrodynamic modes}

\label{sec:hydro}

\subsection{Order parameter}

Consider the following 1-spin operator:\begin{align}
\label{eq:operatorR}
\hat{\mathcal{R}}_{\alpha\beta}=\frac{1}{N}\sum_i \langle \hat{S}_i^{\beta} \rangle \hat{S}_i^{\alpha} .
\end{align}
In both the Edwards-Anderson and correlated spin glass cases we have non-zero expectation values of this operator, $\left\langle \hat{\mathcal{R}}_{\alpha\beta}\right\rangle=q\delta_{\alpha\beta}\neq 0$. For concreteness, the statistical averages are taking with respect to a density matrix operator $\hat{\rho}_{G}$ representing a macrostate $G$, $\langle \cdot \rangle=\textrm{Tr}\left[\hat{\rho}_G \cdot\right]$.

Let us consider now a transformation $g$ of the macrostate, $G\rightarrow G'=gG$, consisting of a global spin rotation, $g\in SO(3)$:\begin{align}
\hat{\rho}_{G'}=\hat{U}\left(\boldsymbol{\theta}\right)\hat{\rho}_{G}\hat{U}^{\dagger}\left(\boldsymbol{\theta}\right), \,\,\textrm{with}\,\,\hat{U}\left(\boldsymbol{\theta}\right)=e^{-i\boldsymbol{\theta}\cdot\hat{\boldsymbol{S}}_{\textrm{tot}}},
\end{align}
and $\hat{\boldsymbol{S}}_{\textrm{tot}}$ is the total spin operator, \begin{align}
\label{eq:operatorS}
\hat{\boldsymbol{S}}_{\textrm{tot}}=\sum_{i}\hat{\boldsymbol{S}}_i.
\end{align}
Note that in a spin glass, $\langle \hat{\boldsymbol{S}}_{\textrm{tot}}\rangle=0$ (no equilibrium magnertization). Generically, $\hat{\rho}_{G'}\neq \hat{\rho}_{G}$; the expectation value of the operator in Eq.~\eqref{eq:operatorR} in the new macrostate $G'$ is \begin{align}
\left\langle\hat{\mathcal{R}}_{\alpha\beta}\right\rangle'=q R_{\alpha\beta}\left(\boldsymbol{\theta}\right),
\end{align}
where $R_{\alpha\beta}(\boldsymbol{\theta})$ are matrix elements of the $SO(3)$ rotation $\hat{R}(\boldsymbol{\theta})$ associated with $\hat{U}(\boldsymbol{\theta})$.

For both the Edwards-Anderson and the correlated spin glass, $G$ and $G'$ represent physically different macrostates but with the same free energy, i.e., rigid spin rotations of the glass introduce no free-energy cost. This is evident for the Edwards-Anderson spin glass of the first model since $\hat{U}(\boldsymbol{\theta})$ is a symmetry of the Hamiltonian in Eq.~\eqref{eq:model1}. The Edwards-Anderson spin glass breaks spontaneously this symmetry, and $\boldsymbol{\theta}=(\theta_x,\theta_y,\theta_z)$ parametrize three branches of soft (Goldstone) modes.\cite{Halperin_Saslow} This is less evident in the case of a correlated spin glass since $\hat{U}(\boldsymbol{\theta})$ is not a symmetry of the Hamiltonian. Yet magnetic anisotropy can be assumed to be completely frustrated in the limit $L_{\textrm{DS}}\gg L_{\textrm{dis}}$. In that case we still expect $\boldsymbol{\theta}$ to parametrize three branches of soft modes dominating the response at low frequencies. However, the dynamics of these modes is not protected by a local conservation law and they should be overdamped at long wavelengths. The low-frequency spin response is diffusive, and that suffices to give a linear in $T$ specific heat ($T\ln1/T$ in $d=2$) at the lowest temperatures.

\subsection{Hydrodynamic operators}

The argument goes as follows. The idea is that the long-time response of the spin glass is dominated by collective modes associated with fluctuations of the three angles $\boldsymbol{\theta}=(\theta_x,\theta_y,\theta_z)$, which relax on long (hydrodymamic) times compared to the rest of modes of the spin dynamics due to the free-energy invariance described above. A minimal hydrodynamic theory should include these variables as well as a coarse-grained spin density associated with the generator of rotations,\begin{align}
\hat{\boldsymbol{s}}\left(\mathbf{r}\right)=\hbar\sum_{i}\hat{\boldsymbol{S}}_i\delta\left(\mathbf{r}-\mathbf{r}_i\right).
\end{align}
Similarly, we can define the local operator\begin{align}
\hat{R}_{\alpha\beta}\left(\mathbf{r}\right)=\frac{1}{qN}\sum_i \langle \hat{S}_i^{\beta} \rangle \hat{S}_i^{\alpha}\delta\left(\mathbf{r}-\mathbf{r}_i\right).
\end{align}
From the spin commutator algebra we just have
\begin{align}
& \frac{-i}{\hbar}\left[\hat{s}_{\alpha}\left(\mathbf{r}_1\right),\hat{R}_{\beta\gamma}\left(\mathbf{r}_2\right)\right]=\epsilon_{\alpha\beta\lambda}\hat{R}_{\lambda\gamma}\left(\mathbf{r}_1\right)\delta\left(\mathbf{r}_1-\mathbf{r}_2\right).
\end{align}
However, these are not good hydrodynamic operators since (even after coarse-graining) $\langle \hat{R}_{\alpha\beta}(\mathbf{r}) \rangle\neq0$ in a glass. Following Halperin and Saslow, we consider instead\begin{align}
\label{eq:theta_op}
\hat{\theta}_{\alpha}\left(\mathbf{r}\right)=\frac{1}{2}\epsilon_{\alpha\beta\gamma}\hat{R}_{\gamma\beta}(\mathbf{r}).
\end{align}
This definition only makes full sense if the expectation value of operator $\hat{R}_{\gamma\beta}(\mathbf{r})$ does not deviate much from the Edwards-Anderson order parameter, $\langle \hat{R}_{\alpha\beta}(\mathbf{r}) \rangle\approx\delta_{\alpha\beta}$, and therefore, $\langle \hat{\theta}_{\alpha}(\mathbf{r}) \rangle\approx 0$ upon appropriate coarse-graining. 

Next, we consider spectral functions of two arbitrary operators $\hat{A}$, $\hat{B}$ with time dependence described in the Heisenberg picture, \begin{align}
\chi_{\hat{A},\hat{B}}''(\omega)=\frac{1}{2\hbar}\int_{-\infty}^{\infty}dt\,e^{i\omega t}\left\langle\left[\hat{A}(t),\hat{B}^{\dagger}(0)\right]  \right\rangle,
\end{align}
and complex response functions given by\begin{align}
\chi_{\hat{A},\hat{B}}(z)=\int_{-\infty}^{\infty}\frac{d\omega}{\pi}\frac{\chi_{\hat{A},\hat{B}}''(\omega)}{\omega-z}\,\, (\textrm{Im} z\neq 0).
\end{align}
As stated before, the low-frequency response of the system is dominated by fluctuations of the Fourier components of operators $\hat{\boldsymbol{\theta}}(\mathbf{r})$, $\hat{\boldsymbol{s}}(\mathbf{r})$; in particular, cross correlations satisfy the sum rule\begin{align}
\label{eq:cross_corr}
 \int_{-\infty}^{\infty} & \frac{d\omega}{\pi}\, \chi_{\hat{\theta}_{\alpha,\mathbf{k}_1},\hat{s}_{\beta,\mathbf{k}_2}}''(\omega)=\frac{1}{\hbar}\left\langle\left[\hat{\theta}_{\alpha,\mathbf{k}_1},\hat{s}_{\beta,-\mathbf{k}_2}\right] \right\rangle \\
& = \frac{i}{2V}\int d^d\mathbf{r}\,e^{-i\left(\mathbf{k}_1-\mathbf{k}_2\right)\cdot\mathbf{r}}\left[\delta_{\alpha\beta}\left\langle \hat{R}_{\gamma\gamma}(\mathbf{r}) \right\rangle - \left\langle \hat{R}_{\alpha\beta}(\mathbf{r}) \right\rangle \right]\nonumber\\
&\approx i\delta_{\alpha\beta}\delta_{\mathbf{k}_1,\mathbf{k}_2}.
\nonumber
\end{align}
The last result is true if $\langle \hat{R}_{\alpha\beta}(\mathbf{r}) \rangle\approx\delta_{\alpha\beta}$. 
It means that if coarse-grained (without worrying much about the exact procedure, only that it exists) density operators associated with rotation angles of the $SO(3)$ order parameter and the generator of rotations can be defined locally, then these operators must be related by canonical conjugation relations quickly decaying in space. There is at least one situation in which this is questionable and will be discussed later. Note that the last result does not guarantee the existence of Goldstone modes, which require the more stringent sum rules\cite{Lange,Forster}\begin{align}
\lim_{\mathbf{k}_1,\mathbf{k}_2\rightarrow 0}\int_{-\infty}^{\infty}\frac{d\omega}{\pi}\,\omega^n\,\chi_{\hat{\theta}_{\alpha,\mathbf{k}_1},\hat{s}_{\beta,\mathbf{k}_2}}''(\omega)=0\,\,\, \forall n>0.
\end{align}
To probe the latter, one needs to show 1) that there are no long-range forces between local operators and 2) that there is a local conservation law for the coarse-grained spin density. So long $J_{ij}$ decays quickly as the separation of sites $i$ and $j$ grows, condition 1) should be satisfied for local operators. Condition 2) can only be the result of a microscopic symmetry, which is present in model 1, but not in model 2.

\subsection{Dispersion relation}

Consider now specifically the complex response functions $\chi_{\hat{\theta}_i,\hat{\theta}_j}(z)$ grouped in a matrix, $\left[\hat{\chi}(z)\right]_{ij}=\chi_{\hat{\theta}_i,\hat{\theta}_j}(z)$, where the subscript $i$ span spin and spatial indices/coordinates; in particular, $\chi_{\hat{\theta}_i,\hat{\theta}_j}=\chi_{\hat{\theta}_{\alpha}(\mathbf{r}_1),\hat{\theta}_{\beta}(\mathbf{r}_2)}$ with $i=(\alpha,\mathbf{r}_1)$, $j=(\alpha,\mathbf{r}_2)$ describes the local rotation of the order parameter in position $\mathbf{r}_1$ due to a magnetic torque produced by e.g. spin accumulation\cite{spin-transfer} in position $\mathbf{r}_2$. For its Fourier components, $i=(\alpha,\mathbf{k}_1)$, $j=(\alpha,\mathbf{k}_2)$, 
the result in Eq.~\eqref{eq:cross_corr} allows us to write a dispersion-relation representation of the form\begin{align}
\label{eq:dispersion_relation_rep}
\hat{\chi}(z)=\left[\hat{\chi}^{-1}(0)-\frac{\hbar^2\chi_{\mathbf{m}}}{g^2\mu_B^2}z^2-iz\,\hat{\Sigma}(z)\right]^{-1}.
\end{align}
The first term is the static susceptibility of rotation variables corresponding to a second derivative of the free energy, $\left[\hat{\chi}^{-1}(0)\right]_{ij}=\partial^2 F/\partial\theta_i\partial\theta_j$. Given the free-energy invariance with respect to spin rotations discussed before, $\hat{\chi}^{-1}(0)$ should go to $0$ in the limit $\mathbf{k}_1,\mathbf{k}_2\rightarrow 0$. We parametrize this by\begin{align}
\left[\hat{\chi}^{-1}(0)\right]_{(\alpha,\mathbf{k}_1),(\beta,\mathbf{k}_2)}=A_{\alpha}|\mathbf{k}_1|^2\delta_{\alpha,\beta}\delta_{\mathbf{k}_1,\mathbf{k}_2},
\end{align}
where $A_{\alpha}$ is some fraction of $Jqa^{2-d}$ and represents the order-parameter stiffness with respect to rotations along $\alpha$-axis. Here I am retaining only diagonal terms in momenta assuming that translational invariance is effectively restored at long wavelengths (or upon coarse-graining of hydrodynamical variables). 

In the second term we have the magnetic susceptibility relating the macroscopic magnetization density with an applied magnetic field, $\mathbf{m}=\mathbf{M}/V=\chi_{\mathbf{m}}\mathbf{B}$. In a spin glass the response is similar to an isotropic paramagnet; in particular, in the correlated spin glass from Imry-Ma arguments one can write\begin{align}
\chi_{\mathbf{m}}\approx \frac{g^2\mu_B^2}{qa^dK}\left(\frac{L_{\textrm{c}}}{L_{\textrm{dis}}}\right)^{\frac{d}{2}}= \frac{g^2\mu_B^2}{qa^dK}\left(\frac{L_{\textrm{DW}}}{L_{\textrm{dis}}}\right)^{\frac{2d}{4-d}}.
\end{align}

Finally, the third term is the Laplace transform of a memory matrix 
describing the effect of magnetic forces produced by the rest of microscopic degrees of freedom on the hydrodynamic variables. In the Mori formalism,\cite{Forster} the memory matrix can be written as a Kubo correlation function of the spin current density operator, $\dot{\hat{\boldsymbol{s}}}=-i[\hat{\boldsymbol{s}},\hat{H}]/\hbar$, with hydrodynamic modes projected-out from the Hamiltonian evolution. The important observation is that for a Hamiltonian of the type of model 1, there must be a local conservation law for the (coarse-grained) spin density as a consequence of the symmetry under spin rotations, implying that $\lim_{\mathbf{k}_1,\mathbf{k}_2\rightarrow 0}\Sigma(z)\rightarrow 0$. In this case, the restoring variables $\boldsymbol{\theta}$ parametrize long-lived excitation with lifetimes satisfying $\frac{1}{\tau}\propto |\mathbf{k}|^2$. These are the Halperin-Saslow waves of spin glasses.\cite{Halperin_Saslow}

However, for microscopic Hamiltonian of the type of model 2, in which anisotropies or other form relativistic corrections to the spin Hamiltonian are present, spin angular momentum is not conserved. In the long-wavelength limit we have \begin{align}
\lim_{\mathbf{k}_1,\mathbf{k}_2\rightarrow 0}\left[\hat{\Sigma}(z)\right]_{(\alpha,\mathbf{k}_1),(\beta,\mathbf{k}_2)}=\frac{1}{V}\int_{-\infty}^{\infty}\frac{d\omega}{i\pi}\frac{\chi_{\hat{\tau}_{\alpha},\hat{\tau}_{\beta}}''\left(\omega\right)}{\omega(\omega-z)},
\end{align}
where $\hat{\boldsymbol{\tau}}\equiv -i\left[\hat{\boldsymbol{S}}_{\textrm{tot}},\hat{H}\right]$ is the spin torque operator. We conclude that in the hydrodynamical limit (taking $z=\omega+i0^{+}\rightarrow 0$) the response in Eq.~\eqref{eq:dispersion_relation_rep} can be approximated by\begin{align}
\label{eq:approx_chi}
\chi_{(\alpha,\mathbf{k}),(\beta,\mathbf{k})}(\omega)\approx\frac{g^2\mu_B^2}{\hbar^2\chi_{\mathbf{m}}}\frac{\delta_{\alpha\beta}}{c_{\alpha}^2|\mathbf{k}|^2-\omega^2-i\gamma_{\alpha}\omega},
\end{align}
where $c_{\alpha}$ is the spin-wave velocity of the correlated spin glass,\begin{align}
c_{\alpha}=\sqrt{\frac{g^2\mu_B^2A_{\alpha}}{\hbar^2\chi_{\mathbf{m}}}}\sim\frac{aq\sqrt{JK}}{\hbar}\left(\frac{L_{\textrm{dis}}}{L_{\textrm{DW}}}\right)^{\frac{d}{4-d}},
\end{align}
and $\gamma_{\alpha}$ is a damping coefficient given by a Green-Kubo formula of the form\begin{align}
\gamma_{\alpha}=\frac{g^2\mu_B^2}{\hbar^2 V\chi_{\mathbf{m}}}\lim_{\omega\rightarrow 0}\frac{2}{\omega}\chi_{\hat{\tau}_{\alpha},\hat{\tau}_{\alpha}}''\left(\omega\right).
\end{align}

\begin{figure}[t!]
\includegraphics[width=\columnwidth]{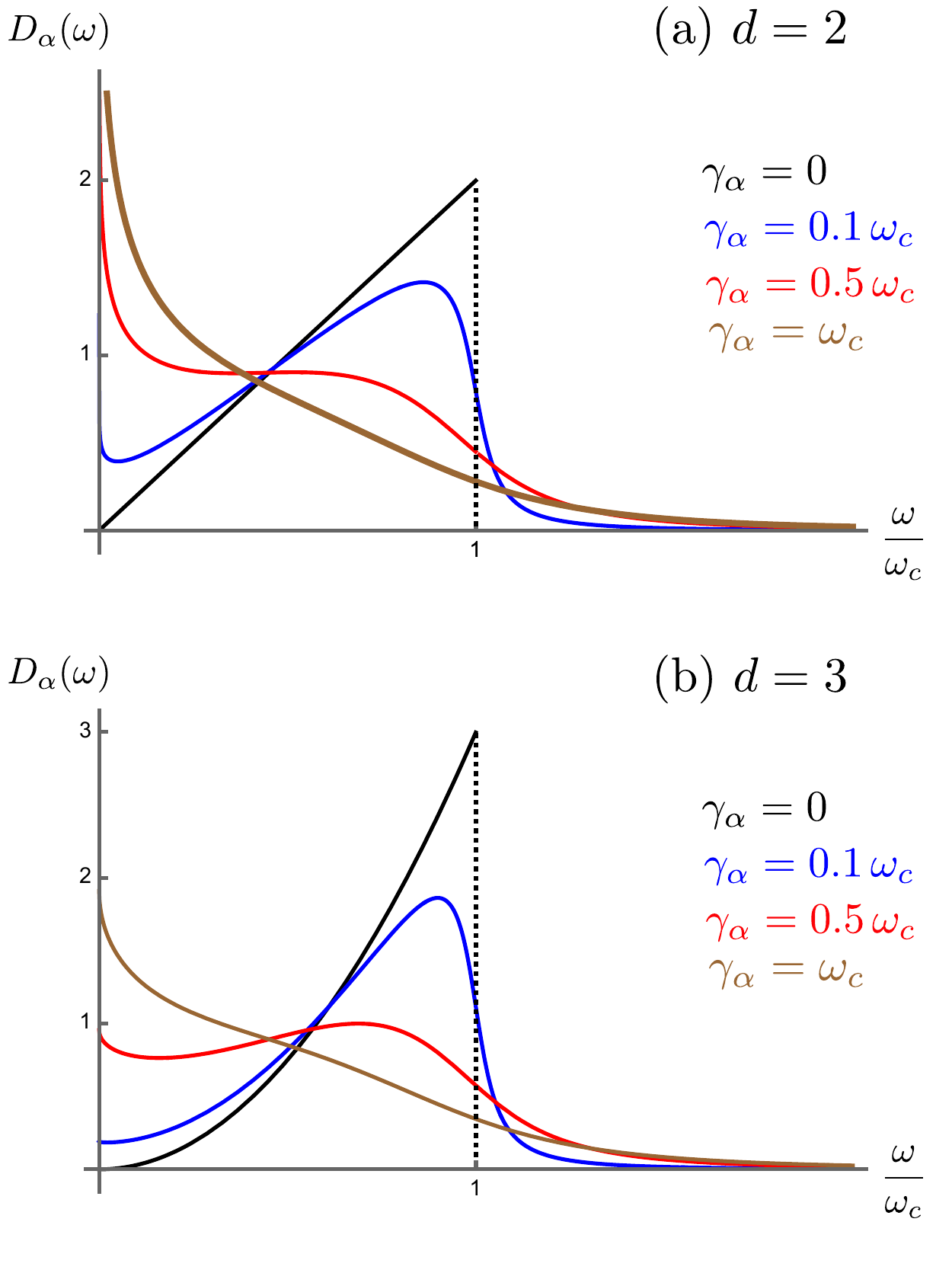}
\caption{\label{fig:DOS} Density of states of branch $\alpha$ in units of the inverse of the cut-off frequency $\omega_c$ as deduced from the numerical evaluation of Eq.~\eqref{eq:DOS} for different values of the damping coefficient $\gamma_{\alpha}$ in (a) $d=2$ and (b) $d=3$ solids.}
\end{figure}

The poles of the susceptibility in Eq.~\eqref{eq:approx_chi} lie at\begin{subequations}
\label{eq:poles}
\begin{align}
\omega_{\alpha}(\mathbf{k})& =\pm\sqrt{c_{\alpha}^2|\mathbf{k}|^2-\frac{i\gamma_{\alpha}^2}{4}}-i\frac{\gamma_{\alpha}}{2}\\
& \approx \begin{cases}
 \pm c_{\alpha}|\mathbf{k}|-i\frac{\gamma_{\alpha}}{2}\,\,\, & \textrm{if}\,\, c_{\alpha}|\mathbf{k}|\gg \frac{\gamma_{\alpha}}{2},\\
 -i\gamma_{\alpha}, -i\frac{ c_{\alpha}^2|\mathbf{k}|^2}{\gamma_{\alpha}}\,\,\, & \textrm{if}\,\, c_{\alpha}|\mathbf{k}|\ll \frac{\gamma_{\alpha}}{2}.
\end{cases}
\end{align}
\end{subequations}
In the first limit we have propagating spin waves of the type of Halpein-Saslow. In the second limit, one evolves into a fully gapped relaxation mode, and the other to a diffusive mode which grabs most of the spectral weight at low frequencies.

\section{Specific heat}

The specific heat per site at constant field (taken $0$) can be written as\begin{align}
C=T\left(\frac{\partial S}{\partial T}\right)_{\mathbf{B}=0}=k_B\int_0^{\infty}d\omega\, \left(\frac{\hbar\omega}{k_B T}\right)^2\sinh^{-2}\left(\frac{\hbar\omega}{k_B T}\right)D\left(\omega\right),
\end{align}
where $D(\omega)$ is the density of modes per site at frequency $\omega$. At low frequencies, we have argued that the response is dominated by the collective modes $\theta_{\alpha}$ in Eq.~\eqref{eq:dispersion_relation_rep}, defining\begin{align}
D\left(\omega\right)=\frac{2}{\pi N\omega}\textrm{Tr}\left[\hat{\chi}^{-1}(0)\cdot\textrm{Im}\hat{\chi}\left(\omega+i0^+\right)\right].
\end{align}
Here the trace is over indices $i=(\alpha,\mathbf{k})$. In the approximation of Eq.~\eqref{eq:approx_chi}, each branch labelled by $\alpha$ contributes as\begin{subequations}
\label{eq:DOS}
\begin{align}
& D_{\alpha}\left(\omega\right)=\omega_c^{-1}f_d\left(\frac{\omega}{\omega_c},\frac{\gamma_{\alpha}}{\omega_c}\right),\,\,\textrm{with}\\
& f_d\left(x,y\right)=\frac{2dy}{\pi}\int_0^1 dz\,\frac{z^{d+1}}{\left(z^2-x^2\right)^2+x^2y^2}.
\end{align}
\end{subequations}
The remaining integral in the definition of the dimensionless function $ f_d\left(x,y\right)$ comes from the summation over momenta. The subindex $d=2,3$ refers to dimensionality of the systems considered below. Note that in the formula for $d=2$ the effects of quantum confinement in thin films\cite{Sidorova_etal} are not taken into account but they could be included following the same steps as in Ref.~\onlinecite{Zaccone}. The specific heat reads\begin{subequations}
\label{eq:C}
\begin{align}
& C=k_B\, g_d\left(\frac{T}{T_c},\frac{T_{\gamma}}{T_c}\right),\,\,\textrm{with}\\
& g_d\left(x,y\right)=2x\int_0^{\infty} dz\,z^2\sinh^{-2}z\, f_d\left(2xz,y\right),
\end{align}
\end{subequations}
and where $T_{\gamma}\equiv\hbar\gamma_{\alpha}/k_B$, $T_{c}\equiv\hbar\omega_c/k_B$. In all these expressions $\omega_c$ is a cut-off for the linear dispersion of the collective modes, which is ultimately related to the inverse of the coarse-graining length of hydrodynamic variables proportional to $L_{\textrm{c}}$ in Eq.~\eqref{eq:Lc}.

If $\gamma_{\alpha}=0$, then $f_d(x,0)=dx^{d-1}$, and at low temperatures one recovers Debye's law for the contribution of mode $\alpha$ to the specific heat, $C\sim T^d$. However, for finite damping, the evaluation of Eq.~\eqref{eq:DOS} in Fig.~\ref{fig:DOS} shows that there is a transfer of spectral weight from high to low energies required by the presence of a broad diffusive peak in the response function at long wavelengths. In $d=3$ the density of states is finite at $\omega=0$, and diverges logarithmically in the case of $d=2$. This gives rise to a deviation of the specific heat at low temperatures $T\ll T_*$ of the form\begin{align}
\label{eq:C_anal}
C=k_B\times\begin{cases}
 \frac{2\pi TT_{\gamma}}{3T_c^2}\ln\left(\frac{T_c^2}{2T_{\gamma}T}\right)\,\,\, & \textrm{if}\,\, d=2,\\
 \frac{2\pi TT_{\gamma}}{T_c^2}\,\,\, & \textrm{if}\,\, d=3.
\end{cases}
\end{align}

\begin{figure}[t!]
\includegraphics[width=\columnwidth]{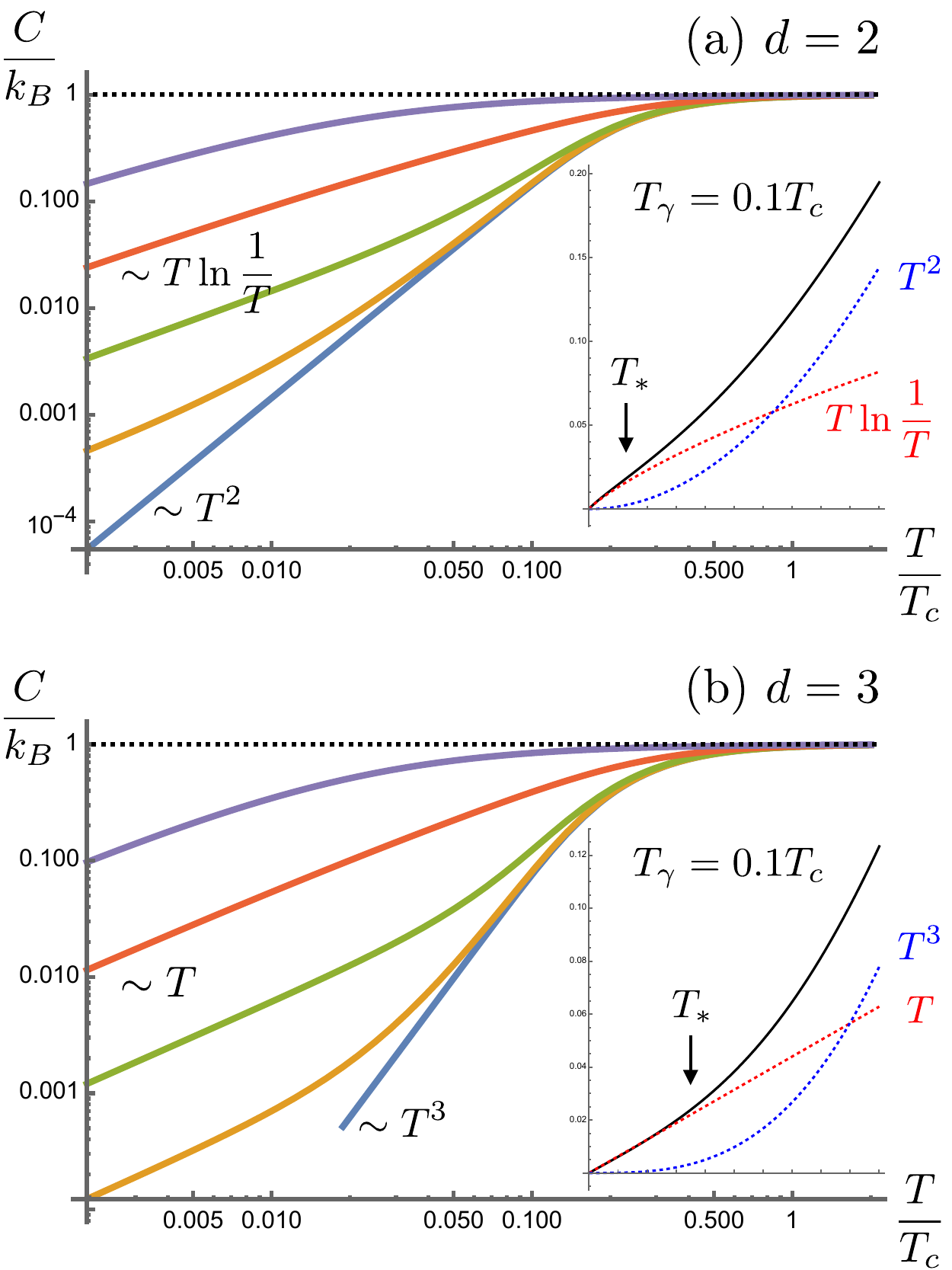}
\caption{\label{fig:C} Specific heat as a function of temperature 
for different values of $\gamma_{\alpha}$ (from bottom to top in units of $\omega_c$: $0$, $0.01$, $0.1$, $1$, $10$). 
The inset shows the specific heat as a function of temperature 
for the lowest temperatures, $T<T_{\gamma}=0.1T_c$. The dashed blue and red lines correspond to the analytical estimate in Eq.~\ref{eq:C_anal} and Debye theory, respectively.}
\end{figure}

What is the crossover temperature $T_{*}$? Figure~\ref{fig:DOS} shows that there are two regimes we must consider. If $\gamma_{\alpha}<2\omega_c$, only a fraction of the spin waves are overdamped. This translates to a non-monotonic behavior of $D_{\alpha}(\omega)$: it first decreases, then it grows following the otherwise expected $\omega^{d-1}$ dependence, and then it decreases again to make up for the excess of density of states at low frequencies. In $d=2$, the logarithmic dependence in $f_2(x,y)$ ceases to be appreciable when $x\sim y/2\pi$. In $d=3$, it is around $x^2\sim 2y/\pi$. However, if $\gamma_{\alpha}>2\omega_c$, then the whole spectrum is incoherent and the density of states is a monotonically decreasing function of $\omega$, fully dominated by the diffusive poles. Moreover, in this regime the density of states scales as $f_{d}(x,y)=yh_d(xy)$, 
i.e., the dependence on the first argument $x$ is really through the combination $xy$, and
thus the separation between high and low temperatures is given by the condition $xy\sim 1$. Putting all these arguments together, one finds\begin{align}
\label{eq:T*}
T_*\approx\begin{cases}
 \textrm{min}\left[\frac{T_{\gamma}}{4\pi},\frac{T_c^2}{2T_{\gamma}}\right]\,\,\, & \textrm{if}\,\, d=2,\\
  \textrm{min}\left[\sqrt{\frac{T_cT_{\gamma}}{2\pi}},\frac{T_c^2}{2T_{\gamma}}\right] \,\,\, & \textrm{if}\,\, d=3.
\end{cases}
\end{align}

Figure~\ref{fig:C} shows $C$ as a function of $T$ in double logarithmic scale from the numerical evaluation of Eq.~\eqref{eq:C}. The calculation agrees well with the analytical estimates in Eq.~\eqref{eq:C_anal} for the lowest temperatures $T< T_*$ estimated in Eq.~\eqref{eq:T*}. The two regimes in the density of states translates to two different scenarios for the temperature dependence of the specific heat. When the density of states is non-monotonic, $C$ displays three different behaviors with $T$: The anomalous dependence in Eq.\eqref{eq:C_anal} (incoherent regime), followed by the otherwise expected $T^d$ dependence (Debye regime) at $T_*<T_c$, and finally the classical result (equipartition regime) around $T_c$. The insets in Fig.~\ref{fig:C} illustrates the first crossover in linear scale. As the damping coefficient increases, $T_*$ approaches $T_c$ and the Debye regime disappears. When the density of states is monotonically decreasing, i.e., when all the modes are overdamped, there is a direct crossover from the incoherent to the equipartition regime at $T_*\ll T_c$.

\section{Disclinations}

One instance in which the definition of local operators $\hat{\theta}_{\alpha}$ is questionable is if the non-collinear texture characterizing the glass state does not admit a description in terms of a smoothly varying rotation of a local spin frame due to the presence of $\mathbb{Z}_2$ vortex disclinations. The possibility of these defects can be understood from the topology of the $SO(3)$ manifold introduced before. Since the $SU(2)$ group of spin rotations is isomorphic to the unit hypersphere $S^3$, a generic operator $\hat{U}$ can be represented by a 4-dimensional unit vector (or quaternion) $q=(w,\boldsymbol{v})$, with $w^2+|\boldsymbol{v}|^2=1$. The elements of the associated $SO(3)$ matrix are\begin{align}
R_{\alpha\beta}=(1-2|\boldsymbol{v}|^2)\delta_{\alpha\beta}+2 v_{\alpha}v_{\beta}-2\epsilon_{\alpha\beta\gamma}.
\end{align}
Since $q$ and $-q$ correspond to the same rotation matrix, we can conclude that $SO(3)\cong \mathbb{RP}^3$, i.e., the hypersphere $S^3$ with antipodal points identified as the same.

The macroscopic description of these defects is provided by a non-abelian generalization of the disclination density in 2D crystals. It is convenient to introduce the \textit{velocity} field $\boldsymbol{\Omega}_{i}=2\partial_iq\wedge q^{*}$, where $q^{*}=(w,-\boldsymbol{v})$ and $\wedge$ represents the Hamilton product of two quaternions (i.e., a matrix product); by definition\cite{foot} the first entry in $\boldsymbol{\Omega}_{i}$ is identically zero (i.e., $\boldsymbol{\Omega}_{i}$ is really a vector in spin space, not a matrix). This field plays a role analogous to the strain tensor. 
The density of disclination lines is given by\begin{align}
\epsilon_{ijk}\boldsymbol{\rho}_k=\partial_i\boldsymbol{\Omega}_j-\partial_j\boldsymbol{\Omega}_i-\boldsymbol{\Omega}_i\times \boldsymbol{\Omega}_j.
\end{align}
Similarly, the disclination current can be defined as\begin{align}
\boldsymbol{j}_i=\dot{\boldsymbol{\Omega}}_i-\partial_i\boldsymbol{\Omega}_0-\boldsymbol{\Omega}_0\times \boldsymbol{\Omega}_i,
\end{align}
where $\boldsymbol{\Omega}_0=2\dot{q}\wedge q^{*}$ is the precession frequency. The continuity equation for disclinations follows from the geometrical relation\begin{align}
D_t\boldsymbol{\rho}_i+\epsilon_{ijk}D_j\boldsymbol{j}_k=0,
\end{align}
which is the Jacobi identity for the covariant derivative $D_{\mu}=\partial_{\mu}-\boldsymbol{\Omega}_{\mu}\times$.

If the quaternion field is single valued, $\partial_{i}\partial_jq=\partial_{j}\partial_iq$, then we have $\boldsymbol{\rho}_k=0$ in the definition above. The property $\partial_i\boldsymbol{\Omega}_j-\partial_j\boldsymbol{\Omega}_i=\boldsymbol{\Omega}_i\times \boldsymbol{\Omega}_j$ is akin to the Mermin-Ho relation\cite{Mermin-Ho} in the A-phase of superfluid $^{3}$He. It expresses that vorticity in one of the angles $\theta_{\alpha}$ parametrizing the local spin frame can be unwinded (in multiples of $4\pi$) by a texture of the Edwards-Anderson order parameter and does not introduce a singularity. On the contrary, if this relation is not satisfied, it implies winding in multiples of $2\pi$, which cannot be undone smoothly: There is a line of vortex cores associated with an essential branch cut where the order parameter is multivalued (where antipodal points of the $SO(3)$ hypersphere are identified).

For a glass described by the Hamiltonian in Eq.~\eqref{eq:model1}, Volovik and Dzyaloshinskii proposed\cite{DV1,DV2} that in account for disclinations the macroscopic spin dynamics (coarse-grained over many separated vortex cores) should be derived from a phase-space Lagrangian density of the form $\mathcal{L}=\frac{\hbar}{g\mu_B}\boldsymbol{\Omega}_0\cdot\mathbf{m}-\mathcal{H}[\mathbf{m},\boldsymbol{\Omega}_i]$, with the Hamiltonian density accounting for the free energy cost of non-equilibrium magnetization and exchange couplings,\begin{align}
\label{eq:H3}
\mathcal{H}[\mathbf{m},\boldsymbol{\Omega}_i]=\frac{\mathbf{m}^2}{2\chi_{\mathbf{m}}}+\frac{A}{2}\boldsymbol{\Omega}_i^2,
\end{align}
and supplemented by Poisson brackets derived from the spin operator algebra,\cite{DV3} \begin{align}
\label{eq:Poisson}
\left\{m_{\alpha}(\mathbf{r}),\Omega_{i}^{\beta}(\mathbf{r}')\right\}=\frac{g\mu_B}{\hbar}\left[-\delta_{\alpha\beta}\partial_i+\epsilon_{\alpha\beta\gamma}\,\Omega_{i}^{\gamma}(\mathbf{r})\right]\delta\left(\mathbf{r}-\mathbf{r}'\right).
\end{align}
The equations of motion derived from this theory are\begin{subequations}\begin{align}
& \frac{\hbar}{g\mu_B}\dot{\mathbf{m}}=A\,\partial_i\boldsymbol{\Omega}_i,\\
& \dot{\boldsymbol{\Omega}}_i=\frac{g\mu_B}{\hbar\chi_{\mathbf{m}}}\left(\partial_i\mathbf{m}+\mathbf{m}\times\boldsymbol{\Omega}_i\right).
\end{align}
\end{subequations}
The first one describes the conservation of spin angular momentum. Together with the constitutive relation $\mathbf{m}=\frac{\hbar\chi_{\mathbf{m}}}{g\mu_B}\boldsymbol{\Omega}_0$, the second one is equivalent to $\boldsymbol{j}_i=0$. Thus in the absence of dissipative terms there is no disclination current and $D_t\boldsymbol{\rho}_i=\dot{\boldsymbol{\rho}}_i-\boldsymbol{\Omega}_{0}\times\boldsymbol{\rho}_i=0$, which describes macroscopic spin precession in the presence of a non-equilibrium magnetization density; these oscillations correspond to Halperin-Saslow waves. However, in the presence of dissipative terms the first equation is still true for the model in Eq.~\eqref{eq:model1}, but now there is also a dissipative disclination current $\boldsymbol{j}_{i}=-\gamma\,\boldsymbol{\Omega}_i$. Halperin-Saslow modes disperse as in Eq.~\eqref{eq:poles} and become overdamped at long wavelengths.\cite{DV1,DV2} This mechanism can be understood in analogy with supercurrent relaxation due to vortex motion in superfluids: The disclination current modifies the constitutive relation equivalent to the Josephson formula, introducing damping in the analogue of the second sound.\cite{Davison_etal} These arguments can be extended to disclination motion in crystalline and amorphous solid.\cite{Baggioli_etal,Baggioli_Gouteraux}

\section{Discussion}

The theory can be extended to other amorphous systems not involving magnetic order. The non-collinear magnetic texture describing the correlated spin glass would be a complex ion mass density in structurally disordered materials. In that case the phases $\theta_{\alpha}$ would represent generalized position coordinates of a group of ions. If phase shifts of the mass density produce no free-energy cost, then we would expect a soft mode for each $\theta_{\alpha}$. However, these phases correspond to complex atomic re-arrangements involving motion of a group of ions with respect to others, and consequently, these modes should become diffusive at long wavelengths due to mechanical friction. The best analogy is with the sliding phason in a bipartite incommensurate lattice.\cite{incom1,incom2,incom3} Although TTLS events are typically ascribed to collective dynamics of group of ions too, note the difference between these two scenarios: In the present theory (as elaborated in Sec.~\ref{sec:hydro}), $\theta_{\alpha}$ parametrize small non-equilibrium deviations in which the system remains within a given (local) minimum basin of the free energy associated with macrostate $G$, while in the TTLS theory groups of ions tunnel between local minima.

Back to the central question: Do we need to invoke two branches of excitations, one propagating (acoustic waves) and the other localized (TTLS) in order to explain a deviation from Debye theory? Or is it enough to consider a single branch of excitations that evolve from propagating to diffusive waves? The honest conclusion is that there is no way to discriminate between these two scenarios if we focus on the specific heat alone. In particular, if we look at the insets of Fig.~\ref{fig:C}, 
the total specific heat (black curve) at low temperatures is very well described by the sum of a Debye (in blue) and a anomalous (Eq.~\ref{eq:C_anal}, in red) components, although the black curve was generated by a model including a single branch of excitations.

It is also instructive to consider the present theory in the context of possible glassy phases of strongly correlated materials such as in the cuprate superconductors or in moir\'e systems. Bashan \textit{et al.}\cite{Berg_Schmalian} have recently proposed that scattering off TTLS events in a metallic glass phase can give rise to non-Fermi liquid behavior. 
Note that the cuprates, which are layered materials (thus effectively $d=2$) display a specific heat $C\sim T\ln(1/T)$, which fits well with the present scenario of a single overdamped branch of excitations. In the context of moir\'e systems, Fernandes and I\cite{phason1} have recently proposed that scattering with phason modes of the moir\'e pattern could explain the strange metal behavior observed in twisted bilayer graphene.\cite{strange1,strange2,Jaoui_etal} Damping of the phason modes is the manifestation of interlayer mechanical friction in this case.\cite{phason2}

Nevertheless, there are important differences between these two examples both at the theoretical and phenomenological levels. In the theory of Bashan \textit{et al.} all the degrees of freedom are (in principle) electronic, while in our theory for moir\'e systems the phasons represent lattice degrees of freedom. Note also that cuprates are both strange and \textit{bad metals},\cite{Emery_ Kivelson} while in twisted bilayer graphene the resistivity saturates in apparent compliance with the Mott-Ioffe-Regel limit.\cite{Jaoui_etal} Finally, the inclusion of two different degrees of freedom (mechanic and electronic) introduces also two scales playing the role of $T_c$ above, the new one (Bloch-Gr\"uneisen scale) associated with electron kinematics in the Fermi surface. In a theory involving overdamped soft modes a disparity between Debye and Bloch-Gr\"uneisen temperatures must be invoked in order to explain the simultaneous observation of anomalous specific heat and linear-in-$T$ resistivity since the latter is just an extension of classical equipartition to low temperatures due to the transfer of spectral weight to a diffusive peak. On the contrary, in a theory involving TTLS of electronic origin, the statistical distribution of these events in energy must be assumed to be very broad.

In conclusion, I have presented a discussion of two possible microscopic mechanisms for the transfer of spectral weight of a collective mode from high to low energies in amorphous magnets. In one case it is just the result of magnetic friction with local anisotropy axes that are globally frustrated. In the other case it is the result of the dissipative motion of disclinations. In both cases one has overdamped Halperin-Saslow modes at long wavelengths, which gives rise to a specific heat $C\sim T$ in bulk $d=3$ solids, and $C\sim T\ln(1/T)$ in $d=2$ layered materials, below some scale $T_*$ identified in Eq.~\eqref{eq:T*}. Both mechanism relies on a finite (albeit probably small) stiffness constant. In that regard, note that signatures of long-range spin signals have been reported in amorphous magnets,\cite{exp} which might be an indication of collective spin transport encoded in the rigid dynamics of a non-collinear order parameter.\cite{spin_hydro}



\begin{acknowledgments}
This work is supported by the Spanish MCI/AEI/FEDER, Grant No. PID2021-128760NB-I00. I would like to acknowledge Ricardo Zarzuela, Yaroslav Tserkovnyak and Rafael M. Fernandes for past collaborations that contributed to consolidate some of these ideas. Special thanks should go to Matteo Baggioli for electronic correspondence that clarified some of these ideas before consolidation.
\end{acknowledgments}

\section*{Data Availability Statement}

All plots are generated by direct numerical evaluation of the expressions provided in the body of the manuscript. Model parameters are indicated in the figure and/or in the figures' caption. If the reader has issues with the integrals I will be happy to share my Mathematica notebook.



\nocite{*}

\begin{thebibliography}{99}
\bibitem{foot}  The Hamilton product of two quaternions is defined as $q_1\wedge q_2=(w_1w_2-\boldsymbol{v}_1\cdot\boldsymbol{v}_2,w_1\boldsymbol{v}_2+w_2\boldsymbol{v}_1+\boldsymbol{v}_1\times\boldsymbol{v}_2)$.
\end{thebibliography}

\end{document}